\documentstyle[aps,prl,multicol,epsf,epsfig]{revtex}
\newcommand{\bcols}{\ifpreprintsty\else\begin{multicols}{2}\fi}
\newcommand{\ecols}{\ifpreprintsty\else\end{multicols}\fi}

\def\s{\sum\limits}
\def\p{\prod\limits}
\def\pa{\partial}
\def\i{\int\limits}
\def\be{\begin{equation}}
\def\e{\end{equation}}
\def\beml{\begin{mathletters}}
\def\eml{\end{mathletters}}
\def\beq{\begin{eqnarray}}
\def\eq{\end{eqnarray}}
\def\ba{\begin{array}}
\def\a{\end{array}}

\def\l{\left}
\def\r{\right}

\def\la{\langle}
\def\ra{\rangle}
\def\det{\,{\rm Det}\,}

\def\ctg{\,{\rm cotan}\,}
\def\im{\,{\rm Im}\,}
\def\re{\,{\rm Re}\,}

\begin{document}

\draft

\date{August 2000}

\title{Signature of wave localisation in the time dependence of
a reflected pulse}

\author{M. Titov$^{1,2}$  and C. W. J. Beenakker$^1$}
\address{{}$^1$Instituut-Lorentz, Universiteit Leiden,
P.\,O.~Box 9506, 2300 RA Leiden,
The Netherlands\\
{}$^2$Petersburg Nuclear Physics Institute, Gatchina, 
188350, Russia
}
\maketitle

\begin{abstract}
The average power spectrum of a pulse reflected by a
disordered medium embedded in an $N$-mode waveguide 
decays in time with a power law $t^{-p}$. We show that the
exponent $p$ increases from $\frac{3}{2}$ to $2$
after $N^2$ scattering times, due to the onset of localisation.
We compare two methods to arrive at this result. 
The first method involves the analytic continuation 
to imaginary absorption rate of a static scattering problem. The 
second method involves the solution of a 
Fokker-Planck equation for the frequency dependent reflection matrix,
by means of a mapping onto a problem in non-Hermitian quantum mechanics.
\end{abstract}


\bcols

The time-dependent amplitude of a wave pulse reflected by 
an inhomogeneous medium
consists of rapid oscillations with a slowly decaying envelope.
The power spectrum $a(\omega,t)$ describes the decay with time $t$
of the envelope of the oscillations with frequency $\omega$. 
It is a basic dynamical observable in optics, acoustics, and seismology \cite{book}.
In the seismological context, the attention has focused 
on randomly layered media, which are a model for the subsurface of the Earth.
The fundamental result of White, Sheng, Zhang, and Papanicolaou \cite{Whi87}
for this problem is that $a(\omega,t)$ decays as $t^{-2}$
for times long compared to the scattering time $\tau_s$ at frequency $\omega$.
The dynamics on this time scale is governed by localisation, since the product
of $\tau_s$ and the wave velocity $c$ equals the localisation length
in one dimension. Although this result for the power spectrum 
is more than a decade old, it has thus far resisted
an extension beyond one-dimensional scattering.

Work towards such an extension by Papanicolaou and coworkers \cite{S,P}
has concentrated on locally layered media, in which the
scattering is one-dimensional on short length scales and three-dimensional
on long length scales. This is most relevant for 
seismological applications. Recent dynamical microwave experiments by Genack
{\it et al.} \cite{Gen99} have motivated us to look at this problem in 
a wave-guide geometry, in which the scattering is fully three-dimensional 
--- but restricted to a finite number $N$ of propagating waveguide modes.
(The single-mode case $N=1$ is statistically equivalent to 
the one-dimensional model of Ref.\ \cite{Whi87}.) 
We find that the long-time decay of the average power spectrum 
is a power law as in the one-dimensional case, but with two exponents:
a decay $\propto t^{-3/2}$ crosses over to a $t^{-2}$ decay 
after a characteristic time $t_c=N^2 \tau_s$. The corresponding characteristic 
length  scale $\sqrt{D t_c}$ (with diffusion constant $D$) is the
localisation length in an $N$-mode waveguide. The crossover
is therefore a dynamical signature of localisation in the
reflectance of a random medium, distinct from the signature
in the transmittance (or conductance) considered 
previously in the literature \cite{AKL,MuzK95,Mirlin}.

Let us first formulate the problem more precisely.
We consider the reflection of a scalar wave (frequency $\omega$) from a
disordered region (length $L$, mean free path $l=c \tau_s$) embedded in 
an $N$-mode waveguide (see Fig.\ 1, inset). We assume that the length $L$ 
is greater than the  localisation length $\xi = N l$, so that transmission 
through the disordered region can be ignored. If in addition the
absorption length is greater than $\xi$, the reflection matrix $r(\omega)$ can be
regarded as unitary. The matrix product 
\be
\label{C}
C(\omega,\delta\omega)=r^{\dagger}(\omega-\case{1}{2}\delta\omega)\
r(\omega+\case{1}{2}\delta\omega)
\e
is unitary for unitary $r$, so that
its eigenvalues are phase factors $\exp(i\phi_{n})$. 

The power spectrum 
for a pulse incident in mode $n$ and detected in mode $m$
is related to $C$ by
\beq
a(\omega, t)&=&\int_{-\infty}^{\infty}d\delta t\, e^{i\omega\delta t}
\la r_{mn}(t)r_{mn}(t+\delta t)\ra\nonumber \\
\label{power}
&=&\frac{1}{N} \int_{-\infty}^{\infty}\frac{d\delta\omega}{2\pi}
e^{-it\delta\omega}
\la{\rm Tr}\, C(\omega, \delta\omega) \ra.
\eq
(We have normalised $\int dt\, a(\omega,t)=1$.)
The phase shifts have the joint distribution 
function $P(\phi_1,\phi_2,\dots,\phi_N)$.
To calculate the average of 
${\rm Tr}\,C$  it suffices to know the
one-point function  $\rho(\phi_1)\!=\! N\!\int d\phi_2
\dots \int d\phi_N\, P(\bbox{\phi})$, since 
$\la{\rm Tr}\, C\ra=\int d\phi\,\rho(\phi)e^{i\phi}$.
We will present two different methods of exact solution.
The first method \cite{comment}
(based on analytic continuation) is simple but
restricted to the one-point function, while the
second method \cite{Whi87} (based on a Fokker-Planck equation) 
is more complicated but gives the entire
distribution function.

Analytic continuation to the imaginary frequency difference 
$\delta\omega=i/\tau_a$ relates $\exp{(i\phi_n)}$ to 
the reflection eigenvalue $R_n$ of an absorbing medium
with absorption time $\tau_a$.
The one-point functions are related by
\be
\label{continuation}
\rho(\phi)=\frac{N}{2\pi}+\frac{1}{\pi}\re{\s_{n=1}^{\infty}
e^{-i n\phi}}\i_0^1 R^n\rho(R)\, dR.
\e
This is a quick and easy way to solve the problem,
since $\rho(R)$
is known exactly as a series of Laguerre polynomials
\cite{Bee96}. The method of analytic continuation 
is restricted to the one-point function because 
averages of negative powers of $\exp{(i\phi_n)}$
are not analytic in the reflection eigenvalues 
\cite{comment}. For example, for 
the two-point function one would need to know
the average $\la\exp{(i\phi_n-i\phi_m)} \ra
\to \la R_n R_m^{-1}\ra$ that diverges in the absorbing 
problem. 

The calculation of the power spectrum from 
Eqs.\ (\ref{power}) and (\ref{continuation})
is easiest in the absence of time-reversal 
symmetry, because $\rho(R)$ then has a 
particularly simple form \cite{Bee96}.
We obtain the power spectrum 
\begin{mathletters}
\label{agen}
\beq
a(\omega,t)&=&-N^{-1}\theta(t)\,
\frac{d}{dt}F\!\l( \frac{t}{2N\alpha\tau_s} \r),\\
F(t)&=&\frac{1}{t+1}\s_{n=0}^{N-1}
\l(\frac{t-1}{t+1} \r)^n
P_n\!\!\l(\frac{t^2+1}{t^2-1}\r),
\eq
\end{mathletters}
where $P_n$ is a Legendre polynomial and $\theta(t)$ is the 
unit step function. The coefficient $\alpha=2,\pi^2,8/3$ for dimensionality 
$d=1,2,3$. 
In the single-mode case Eq.\ (\ref{agen})
simplifies to 
\be
\label{alocalised}
a(\omega,t)=4\tau_s(\omega)[t+4\tau_s(\omega)]^{-2}\theta(t),
\e
which is the result of White {\it et al.} \cite{Whi87}.
It decays as $t^{-2}$. For $N\to \infty$ Eq.\ (\ref{agen})
simplifies to 
\be
\label{adiffusive}
a(\omega,t)=t^{-1}\exp{[-t/ \alpha\tau_s(\omega)]}I_1[t/\alpha\tau_s(\omega)]
\theta(t),
\e
where $I_1$ is a modified Bessel function.
The power spectrum now decays as $t^{-3/2}$. 
For any finite $N$ we find a crossover 
from $a=\sqrt{\alpha\tau_s/2\pi}\,t^{-3/2}$ 
for $\tau_s\ll t\ll N^2 \tau_s$ to
$a=2 N \alpha \tau_s\, t^{-2}$ for $t\gg N^2 \tau_s$. 

In the presence of time-reversal symmetry the
exact expression for $a(\omega,t)$
is more cumbersome but the asymptotics
carries over with minor modifications.
In particular, the large-$N$ limit 
(\ref{adiffusive}) with its $t^{-3/2}$
decay remains the same, while 
the $t^{-2}$ decay changes only in the prefactor:
$a=(N+1)\alpha\tau_s\, t^{-2}$ for $t\gg N^2\tau_s$.

We now turn to the second method of solution,
based on a Fokker-Planck equation for the entire
distribution function
$P(\bbox{\phi})$. The equation for $N=1$
was derived in Ref.\ \cite{Whi87}. The multi-mode
generalisation can be obtained most directly
by analytic continuation of the Fokker-Planck equation
for the probability distribution of the 
reflection eigenvalues --- which is known \cite{Bee96}.
The resulting Fokker-Planck equation for the phase shifts
takes a simple form in the variable $z=\ln\ctg{(\phi/4)}\in
(-\infty,\infty)$ for $\phi\in (0,2\pi)$.
It reads
\beml
\label{FPz}
\beq
&&\s_{n=1}^N\frac{\pa}{\pa z_n}
\l(\frac{\pa P}{\pa z_n}-P\frac{\pa \Omega}{\pa z_n}\r)=0,\\
&&\Omega(\bbox{z})=\s_{n=1}^N\l(
\ln\cosh{z_n}-c_\beta \Delta\sinh{z_n}\r)\nonumber \\
&& \qquad \;\,+\,\,\beta\!\s_{n>m}\ln\l|
\sinh{z_m}-\sinh{z_n}\r|.
\eq
\eml
We defined the dimensionless frequency
increment $\Delta=\alpha\tau_s\delta\omega$ and
abbreviated $c_\beta=\case{1}{2}\beta (N-1)+1$.
The index $\beta=1(2)$ in the presence (absence)
of time-reversal symmetry.
We emphasise that, although the Fokker-Planck equation can be obtained
by analytic continuation, its solution cannot.
Indeed, this would give the solution   
$P\propto e^{\Omega}$, that  fails because it is 
not normalisable.

We proceed as in Ref.\cite{BeeRej94}
by substituting  
$
P({\bbox z})= \Psi({\bbox z}) \exp{
\l[\case{1}{2}\Omega(\bbox{z})\r]},
$
in order to transform the Fokker-Planck equation (\ref{FPz})
into the Schr\"odinger equation
\beml
\label{main}
\beq
&&\s_{n=1}^N\Big[\! -\!\frac{\pa^2}{\pa z^2_n}+
V(z_n)
+
\s_{m\ne n} \! U(z_n,z_m)\, \Big]
\Psi = 0, \\
&&V(z)=c_\beta^2 \l(
\frac{\Delta^2}{4}\cosh^2{z}-\Delta \sinh{z}
\r)  
+\frac{1}{4\cosh^2{z}}+ V_0, \nonumber \\
&&  
U(z,z')=\frac{1}{2}\beta(\beta-2)\,\frac{\cosh^2z+\cosh^2z'}
{\l(\sinh{z}-\sinh{z'}\r)^2}.
\eq
\eml
Here $V_0=\case{1}{12}\beta^2 (N- 1)
(N- 2+6/\beta)+\case{1}{4}$.
By restricting ourselves to $\beta=2$,
the interaction term $U$ vanishes 
and Eq.\ (\ref{main}) has the form of a one-dimensional 
free-fermion problem. 
The general solution is given by the Slater determinant
\be
\label{Slater}
\Psi({\bbox z})=
\det{ \l\{ \psi_{\mu_n}(z_m) \r\}_{n,m=1}^N},
\e
where $\psi_{\mu}(z)$ is an eigenfunction with eigenvalue
$\mu$ of the single-particle equation
\be
\label{Schrodinger}
-\psi_{\mu}''+V\psi_{\mu}=\mu\psi_{\mu}.
\e
The choice of the eigenfunctions
is restricted by the condition
$\sum_{n=1}^{N}\mu_n=0$
that the total eigenvalue vanishes.

We are now faced with an impasse: The Schr\"odinger 
equation (\ref{Schrodinger}) has a real spectrum consisting of
bound states in the potential well $V(z)$.
The bottom of the well is positive for 
sufficiently large $\Delta$, so that the real spectrum contains
only positive $\mu_n$'s. How then are we to satisfy
the condition of zero sum of
the eigenvalues? The way out of this impasse is to allow for
{\em complex} eigenvalues. The corresponding eigenfunctions will
not be square integrable, but that is not a problem 
as long as the probability distribution 
$P({\bbox z})$ remains normalisable. This is a new twist
to the active field of non-Hermitian quantum mechanics
\cite{Hatano}.

The differential equation (\ref{Schrodinger})
is known as the confluent Heun equation \cite{Heun},
but we have found no mention of the complex spectrum 
in the mathematical physics literature --- perhaps because
it was considered unphysical. The complex spectrum 
is constructed by means of a complete set of polynomials 
to order $N-1$,
\be
\label{Ag}
{\cal A}_{\mu}(x)=\s_{m=1}^{N}g_m\,
(x-i)^{m-1}(x+i)^{N-m}.
\e
The vector of coefficients ${\bbox g}=\{g_1,g_2,\dots,g_N\}$
is an eigenvector with eigenvalue $\mu$ of the
$N\times N $
tri-diagonal matrix $M$, with non-zero elements
\beq
\label{grecc}
&&{M}_{nn}=2i\Delta N \!\l(n\! -\!\frac{N+1}{2}\r)
\! +\! 2\!\l(n\! -\!\frac{N+1}{2}\r)^2\! -\frac{N^2-1}{6}, 
\nonumber \\ 
&&M_{n,n+1}=n^2, \qquad M_{n+1,n}=(N-n)^2.
\eq
Since the trace of $M$ is zero, the condition
$\sum_{n}\mu_n=0$ is automatically satisfied.

The complex spectrum of Eq.\ (\ref{Schrodinger})
consists of the eigenvalues $\mu_n$ with
two linearly independent sets of eigenfunctions,
\begin{mathletters}
\label{III}
\beq
\psi^I_{\mu}(z)&=&\sqrt{\cosh{z}} 
\exp{\l(-\case{1}{2}N\Delta\sinh{z}\r)}
{\cal A}_{\mu}(\sinh{z}),\\
\psi^{II}_{\mu}(z)&=&\sqrt{\cosh{z}}
\exp{\l(\case{1}{2}N\Delta\sinh{z}\r)}
{\cal B}_{\mu}(\sinh{z}).
\eq
\end{mathletters}
The functions ${\cal B}_{\mu}$ 
are related to the polynomials ${\cal A}_{\mu}$ by
\be
\label{B}
{\cal B}_{\mu}(x)=
{\cal A}_{\mu}(x)\i_{0}^{\infty}\frac{e^{-N\Delta\,
\displaystyle{x'}}}{(x-x')^2+1}
{\cal A}_{\mu}^{-2}(x-x')\ dx'.
\e
The functions ${\cal A}_{\mu}$ and  ${\cal B}_{\mu}$
form a bi-orthogonal set on the real axis.
We choose the normalisation such that
\be
\label{orthogonality}
\int_{-\infty}^{\infty}dx\ {\cal A}_{\mu_n}(x)
{\cal B}_{\mu_m}(x)=\delta_{nm}.
\e

The solution $\psi^{I}$ of the first kind can not be used
because the resulting distribution $P({\bbox z})$ is not
normalisable. In fact, since 
\be
\label{Vandermonde}
\det\l\{{\cal A}_{\mu_n}(x_m)\r\}_{n,m=1}^N
\propto \p_{n< m}(x_n-x_m),
\e
one sees that the substitution of $\psi^{I}$ into
Eq.\ (\ref{Slater}) yields  the solution
$P\propto e^\Omega$
that we had rejected earlier.
The solution $\psi^{II}$ of the second kind
does give a normalisable distribution,
\beq
P({\bbox z})  \propto  \p_{n< m}(\sinh{z_n}-\sinh{z_m})
\p_{i=1}^N\cosh{z_i}&& \nonumber \\
\times\mbox{ }
\det{ \l\{ {\cal B}_{\mu_{n}}(\sinh{z_m}) \r\}_{n,m=1}^N},&&
\eq
or, in terms of the variable $x=\sinh{z}=\ctg{(\phi/2)}$,
\be
\label{AB}
P({\bbox x})\propto \det
\{ {\cal A}_{\mu_n}(x_m) \}_{n,m=1}^N\
\det
\{ {\cal B}_{\mu_k}(x_l) \}_{k,l=1}^N.
\e
This is the exact solution of Eq.\ (\ref{FPz})
for $\beta=2$. 
Correlation functions of
arbitrary order can be obtained from 
Eq.\ (\ref{AB}) in terms of a series of the
bi-orthogonal functions ${\cal A}_{\mu}$
and ${\cal B}_{\mu}$ \cite{M,F}.
For example, the density of eigenphases
$\rho(\phi)$ is given by
\be
\label{dens1}
\rho(\phi)=\frac{dx}{d\phi}\s_{n=1}^N
{\cal A}_{\mu_n}(x){\cal B}_{\mu_n}(x).
\e

Let us examine this solution more closely
in various limits. For $N=1$ one has
$\mu=0$, ${\cal A}_{\mu}(x)={\rm constant}$,
and we reproduce the known
single-mode result \cite{Whi87,Gorkov}
\begin{equation}
P(\phi)=\frac{2}{\pi}\tau_{s}\delta\omega(1-\cos\phi)^{-1}{\rm
Im}\,e^{\zeta}{\rm Ei}\,(-\zeta), \label{PphiNis1}.
\end{equation}
Here $\zeta=4i\tau_{s}\delta\omega(1-e^{i\phi})^{-1}$ and Ei is the
exponential-integral function. 
For $N>1$ the eigenvalues $\mu_n$
remain real for $\Delta N^2\ll 1$.
In this regime the integral (\ref{B}) is 
easily evaluated,
because one can substitute effectively 
$[(x-x')^2+1]^{-1}\to\pi\delta(x-x')$.
Hence 
${\cal B}_{\mu}(x)\propto \exp{(-N\Delta\, x)}
{\cal A}_{\mu}(x)\, \theta(x)$.
Using again Eq.\ (\ref{Vandermonde}), we obtain
\be
\label{Laguerre}
P({\bbox x})\propto\p_{n<m}(x_n-x_m)^2 \p_{i=1}^{N}
e^{-N\Delta\, \displaystyle{x_i}}\,
\theta(x_i).
\e 
This is the Laguerre ensemble of random-matrix theory. 
The distribution is dominated by $x\!=\!\ctg{(\phi/2)}\gg 1$, so 
that one can replace
$x_n\to 2/\phi_n$ and recover the result \cite{Bee00}
that the inverse time delays,
$1/\tau_n\equiv\lim_{\delta\omega\to 0}\delta\omega/\phi_n$,
are distributed according to the Laguerre ensemble.
The condition $\Delta N^2\ll 1$ for Laguerre statistics
means that the 
characteristic length $L_{\delta\omega}=\sqrt{D/\delta\omega}$
associated with the frequency increment $\delta\omega$
is greater 
than the localisation length $\xi$.
We therefore refer to the regime of validity of 
Eq.\ (\ref{Laguerre}) as the localised regime. 

At the opposite extreme we have  the ballistic regime
$L_{\delta\omega}\ll l$, or
$\Delta\gg 1$. The integral (\ref{B})
is now dominated by $x'\ll 1$, hence
${\cal B}_{\mu}(x)\propto (x^2+1)^{-1}
{\cal A}_{\mu}^{-1}(x)$.  Moreover,
the off-diagonal elements of the
matrix $M$ may be neglected so that
the polynomials have a simple structure:
${\cal A}_{\mu_n}(x)\propto (x+i)^{N-n}(x-i)^{n-1}$.
The corresponding functions ${\cal B}$ are given by
${\cal B}_{\mu_n}(x)\!=\!
\sin^{N+1}{(\phi/2)} \exp{[i\phi(n-N-\case{1}{2})]}$.
The resulting distribution
of the eigenphases in the ballistic regime 
is
\be
\label{circular}
P({\bbox \phi}) \propto\p_{n<m} \l|
e^{i\phi_n}-e^{i\phi_m} \r|^2,
\e
which we recognize as the circular ensemble of random-matrix 
theory \cite{Mehta}. This is as expected, since for large $\delta\omega$
the matrix $C$ is the product of two independent 
reflection matrices $r$,
each of which is uniformly distributed in the unitary group.
The circular ensemble is the corresponding distribution of the
eigenphases. 

\begin{figure}
\epsfig{file=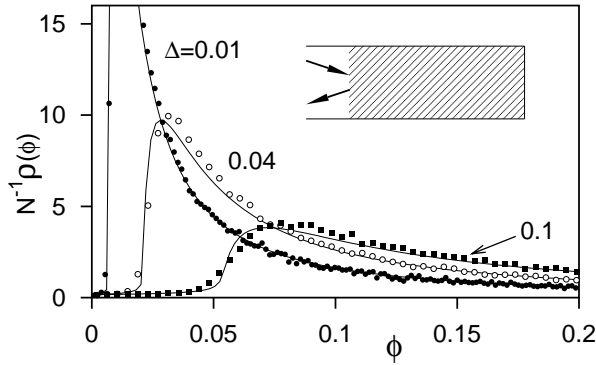,width=8cm}
\vspace*{0.01cm}
\caption{%
Density of the eigenphases for different values of the
dimensionless frequency difference $\Delta=\alpha\tau_s\delta\omega$.
The solid curves are computed from Eq.\ (\ref{R1phi}), the data points
result from a numerical solution of the wave equation
on a two-dimensional square lattice 
($\alpha=\pi^2/4, N=20$; the scattering time $\tau_s$ was obtained
independently from the localisation length). The inset shows the geometry
of a random medium (shaded) embedded in a waveguide.   
}
\label{fig:fphi}
\end{figure}

The intermediate regime  $l\ll L_{\delta\omega}\ll \xi$,
or $N^{-2}\ll \Delta\ll 1$, is the diffusive one.
To study this regime 
we make a WKB approximation of the Schr\"odinger 
equation (\ref{Schrodinger}). 
This approximation requires
$N^2\Delta\gg 1$ and $N\gg 1$, 
hence it contains both the ballistic and the
diffusive regimes.
We obtain
\be
\label{WKB}
\psi^{I,II}_{\mu}(z)=c(\mu)
\l(V(z)-\mu\r)^{-1/4}      
e^{\pm \int_{0}^{z}du\ \sqrt{V(u)-\mu}},
\e
where $c(\mu)$ is a normalisation coefficient.
The ``$+$'' sign in the exponent refers to $\psi_{\mu}^{II}$
and the ``$-$'' sign to $\psi_{\mu}^I$.
The eigenvalues $\mu_n$ densely fill a curve
${\cal C}$ in the complex plane. We may substitute
$\sum_{\mu}f_{\mu}\to \int_{\cal C} \rho(\mu) f(\mu)\, d\mu$,
where $\rho(\mu)=N [4 \pi i c^2(\mu)]^{-1}$ is the eigenvalue density.
For analytic $f(\mu)$, the integral along ${\cal C}$ 
depends only on the end points $\mu_{\pm}=N^2(\frac{1}{3}\pm i\Delta)$
of the curve. From Eqs.\ (\ref{III}), (\ref{dens1}) and (\ref{WKB})
we obtain the eigenphase density
\be
\label{R1phi}
\rho(\phi)=\frac{N}{4\pi\sin^2{(\phi/2)}}
\im\sqrt{\Delta^2+2 i \Delta(1-e^{-i\phi})}.
\e
This result can be obtained also from Eq.\ (\ref{continuation})
in the limit $N\to \infty$. It is derived here
for $\beta=2$, but is actually $\beta$-independent.
(The $\beta$-dependent corrections in the diffusive regime
are due to 
weak localisation, and are smaller by a factor $1/N$.)
One can check that $\rho(\phi)\to N/{2\pi}$ for
$\Delta\gg 1$, as expected in the ballistic regime.
In the opposite regime $\Delta\ll 1$
it simplifies to
\be 
\rho(\phi)=\frac{N\sqrt{\ctg{(\phi/4)}}}
{4\pi \sin{(\phi/2)}}\sqrt{2\Delta-\frac{\Delta^2}{2 \sin{(\phi/2)}}}, 
\e
with the additional restriction: $\sin{(\phi/2)}\ge\Delta/4$. 
We have plotted Eq.\ (\ref{R1phi}) in Fig.\ \ref{fig:fphi} for several
values of $\Delta$.

In conclusion, we have presented 
a signature of localisation in the decay of the power spectrum
of a pulse reflected by a disordered waveguide.
This result is an application of the distribution
of the correlator of the reflection matrix at two different frequencies,
that we have calculated for arbitrary number of modes $N$,
scattering time $\tau_s$, and frequency difference $\delta\omega$.
With increasing $\delta\omega$ the distribution crosses over 
from the Laguerre ensemble in the localised regime
($\delta\omega\ll 1/N^2\tau_s$) 
to the circular ensemble in the ballistic regime
($\delta\omega\gg 1/\tau_s$), 
via an intermediate ``diffusive''
regime. The distribution in this intermediate regime
does not have the form 
of  any of the ensembles known from random-matrix theory
and deserves further study.

This work grew out of an initial investigation
of the single-mode case with K. J. H. van Bemmel and P. W. Brouwer.
We thank H. Schomerus for valuable discussions. 
Our research was supported by the Dutch Science 
Foundation NWO/FOM and by the INTAS grant 97-1342.

\ecols
\end{document}